\documentclass[aps,prd,11pt,amsmath,floats,floatfix, ,superscriptaddress,nofootinbib,showpacs,longbibliography]{revtex4-2}

\usepackage{amsmath, amssymb, graphicx}
\usepackage[normalem]{ulem}
\usepackage[pagebackref=false, colorlinks=true]{hyperref}
\hypersetup{colorlinks,linkcolor=blue,urlcolor=blue}
\usepackage{mathrsfs}
\begin{document}
\preprint{}

\title{Entanglement Entropy for coupled Inverted Harmonic Oscillators}

\author{Srijit Bhattacharjee}
\email{srijuster@gmail.com}
\affiliation{Indian Institute of Information Technology Allahabad (IIITA), Devghat, Jhalwa, Prayagraj-211015, Uttar Pradesh, India.}


\begin{abstract}
The entanglement characteristics of unstable systems in quantum mechanics have been
relatively under-researched. In this note, we study the entanglement entropy for the one-
dimensional coupled inverted harmonic oscillators. Due to the presence of an inverted
parabolic potential, this system possesses runaway solutions. By promoting the eigenfunc-
tions of the Hamiltonian to be members of a Rigged Hilbert space, we find the entanglement
entropy in biorthogonal formulation for a chain of inverted oscillators. The obtained results
are compared with those of harmonic chains. 

\end{abstract}

\maketitle
\section{Introduction}
Harmonic oscillators (HO) play a very crucial role in developing many theoretical models describing physical systems. We have seen intriguing applications of harmonic oscillators in describing quantum systems with many degrees of freedom. Studying HO and their various modified versions would always be important for deciphering fundamental physics. While HO are important ingredients of many branches of physics, the inverted harmonic oscillators (IHOs) also emerge in many interesting situations of physics. It finds applications in quantum tunnelling and decay processes, in quantum Hall systems, in quantum optics, for models demonstrating quantum chaos,  quasinormal modes of black holes, etc \cite{Subramanyan:2020fmx, Hegde:2018xub, Bhattacharyya:2020art, PhysRevD.105.126015}.

Entanglement entropy (EE) is a well-established notion in quantum information theory. Entanglement is now the subject of extensive research in a variety of disciplines, including quantum information, quantum many-body, and black-hole physics. One important finding is that the von Neumann entropy of subregions in the ground state of many-body Hamiltonians with local interactions increases with the subregion's surface area rather than its volume \cite{PhysRevA.66.042327}. See \cite{Eisert:2008ur}
for a comprehensive review on the recent developments. The EE for a subsystem of a system of N-coupled harmonic oscillators demonstrated an area-law-like scaling \cite{Bombelli:1986rw, Srednicki:1993im}, mimicking the second law of black hole thermodynamics \cite{PhysRevD.7.2333, Hawking:1975vcx}. Mostly, these findings are obtained by studying Hermitian systems. Instabilities can also arise in Hermitian Hamiltonian systems through non-standard parameter choices, leading to imaginary or complex frequencies \cite{PhysRevA.89.042312}. The possible extensions of these considerations for non-Hermitian systems have only recently been initiated \cite{Cipolloni:2022fej}. IHOs may offer a useful testing model for developing entanglement and related issues for non-Hermitian systems. IHOs have also been studied extensively in different approaches. Although the Hamiltonian of IHOs can be cast in a Hermitian form, it is not self-adjoint due to the unboundedness of the potential \cite{BARTON1986322, 2000NCimB.115..325S, 2000PhLA..273...37S}. The energy eigenvalues of IHOs are imaginary, posing crucial bottlenecks of developing a meaningful statistical description.   


 In this note we explore IHOs  to shed some light on entanglement and related entropies for such \textit{quasi-hermitian} systems. Throughout the text we set $\hbar=1.$




\section{Inverted Harmonic Oscillator}

In this section, we first review the salient points of the eigenspectra of an inverted harmonic oscillator in one dimension. Unlike a regular oscillator, the IHO describes an inverted potential, representing unstable, runaway solutions (exponential growth/decay) rather than stable oscillations. It should be analysed with sufficient care. The Hamiltonian is given by 
\begin{equation}\label{Hiho}
H = \frac{1}{2}  p^2 
- \frac{1}{2} \omega_{io}^2 x^2.
\end{equation}
This system can formally be obtained from the ordinary harmonic oscillator through the substitution $
\omega \rightarrow i\omega_{io} .
$ The corresponding Schr\"odinger equation becomes
\begin{equation}\label{eigeneq}
\frac{d^2X_E^\pm}{dx^2}
+
(\omega_{io}^2x^2+2E)X_E^\pm=0,
\end{equation}
whose solutions are given by parabolic cylinder functions. The spectrum is doubly degenerate and consists of generalized scattering states normalized through Dirac delta normalization:
\begin{equation}\label{normaliztn}
\int \overline{X_E^\pm(x)}X_{E'}^\pm(x)\,dx
=
\delta(E-E').
\end{equation}

The conventional Hilbert space is not sufficient since the eigenfunctions are not square-integrable. However, incorporating the Rigged Hilbert space (Gel'fand's triplet) structure, one can provide meaningful structures to it. Gel'fand's triplet is represented by
\begin{equation}
S(\mathbb{R})
\subset
L^2(\mathbb{R})
\subset
S^\times(\mathbb{R}),
\end{equation}
where,  $S(\mathbb{R})$ is the Schwartz space,
     $L^2(\mathbb{R})$ is the usual Hilbert space, and
    $S^\times(\mathbb{R})$ is the dual space of tempered distributions (generalized states). For more details, see \cite{Chruscinski:2003uar, PhysRevD.109.105013}.

The IHO eigenstates belong to the dual space $S^\times(\mathbb{R})$, similarly to plane-wave momentum eigenstates in scattering theory. It has to be mentioned here that there exists another set of eigenstates, (solution to \eqref{eigeneq}) obtained by conjugating $\overline{X_E^\pm}=Y_E^\pm.$ 

\subsection{Resonant States}

The eigenvalues are related to the poles of $\overline{X_E^\pm}$ when they are extended in the complex energy plane. The generalized eigenstates $f_n^\pm(x)=\langle x|f_n^\pm\rangle$, are proportional to the residues of these poles: $Res(X^\pm(x);{E_n})\sim f_n^+(x),$ and can be expressed by the complex Hermite polynomials \cite{PhysRevD.109.105013, Chruscinski:2003uar}:
\begin{equation}\label{Hermite}
    f_n^\pm(x)= N_n^\pm H_n(\sqrt{\pm i\omega_{io}}x)e^{\mp i \omega_{io} x^2/2}, 
\end{equation} with $N_n^\pm$ is the n-th normalization constant, usually a complex number.
  These eigenstates satisfy the following properties:

\begin{enumerate}
    \item[i.] They are conjugated to each other:
    \begin{equation}
    f_n^\pm(x)
    =
    \overline{f_n^\mp(x)}.
    \end{equation}

    \item[ii.] They satisfy the biorthogonality relation:
    \begin{equation}
    \langle f_n^\pm | f_m^\mp \rangle
    =
    \delta_{nm}.
    \end{equation}

    \item[iii.] They satisfy the completeness relations:
    \begin{equation} 
    \sum_{n=0}^{\infty}
    |f_n^\pm\rangle
    \langle f_n^\mp|=\mathbb{I}
    ,\end{equation} and
\begin{equation} 
    \sum_{n=0}^{\infty}
    \overline{f_n^\pm(x)}
   f_n^\mp(y)=\delta(x-y).
   \label{deltafn} \end{equation}
\end{enumerate}

 \footnote{The relation in Eq. (\ref{deltafn}) can be obtained using the orthonormality properties of the Hermite polynomials.}These generalized eigenvectors give rise to imaginary eigenvalues $\mp in(+1/2)\omega_{io}$ for an IHO. The eigenvectors are to be understood as tempered distributions in  $S^\times(\mathbb{R})$. The dual of Schwarz space $S^\times(\mathbb{R})$ is decomposed into two (Hardy) sectors accommodating the two sets of eigenstates:
$
|f_n^+\rangle \in S_+,
|f_n^-\rangle \in S_-. $ 
The generalized eigenstates exhibit an intrinsic time asymmetry. They can be demonstrated via the two time evolutions:
\begin{equation}
U_+(t): t\le0,
\qquad
U_-(t): t\ge0.
\end{equation}  The states
\(
f_n^{+}\in S_{-}(\mathbb{R}_x)^{\times}
\)
exist only for forward time evolution (\(t\geq0\)), whereas
\(
f_n^{-}\in S_{+}(\mathbb{R}_x)^{\times}
\)
exist only for backward time evolution (\(t\leq0\)). Consequently, the states
\begin{equation}
f_n^{+}(t,x)=U_{+}(t)\,f_n^{+}(x),
\qquad t\geq0,
\end{equation}
are referred to as \emph{decay states}, while
\begin{equation}
f_n^{-}(t,x)=U_{-}(t)\,f_n^{-}(x),
\qquad t\leq0,
\end{equation}
are known as \emph{growth states}.The eigenstates can be related to the each other through a time reflection symmetry $T|f_n^+\pm \rangle =|f_n^\mp \rangle.$ These eigenstates are sometimes called resonant states, and their related physical energy distributions produce Breit-Wigner resonanance formula. The half-widths or decay widths $\Gamma_n^\pm=(n+1/2)\omega_{io}$ are quantized for IHO. The {\it{ground states}} or \textit{zeroth states} are the most stable in terms of their mean lifetime \cite{2000NCimB.115..325S}. 

\section{Entanglement Entropy for Coupled Inverted Harmonic Oscillators}

Before considering a chain of IHO, let us first recall the method of obtaining the entanglement entropy for a harmonic chain briefly. The procedure closely follows \cite{Srednicki:1993im}, and we now review salient features of it. Consider a system of $N$ coupled harmonic oscillators:
\begin{align}
H &= \frac{1}{2} \sum_{i=1}^{N} p_i^2 
+ \frac{1}{2} \sum_{i,j=1}^{N} x_i K_{ij} x_j\nonumber \\&=\frac{1}{2} \sum_{i=1}^{N} (p_i^2 +\omega^2 x_i^2)
+ {k\over 2}  \sum_{i=1}^{N-1}(x_i-x_{i+1})^2
\end{align}
where $K_{ij}$ is a real, symmetric, positive-definite matrix. For simplicity, we consider only nearest-neighbour couplings-
\[K_{ij}=(\omega^2+k) \delta_{ij}-k(\delta_{i,j+1}+\delta_{i,j-1}).\]

 We now divide our composite system $A\cup B$ into two
subsystems $A,B$ consisting, respectively, of the first $n$ oscillators and the remaining $N-n$ oscillators with $n < N-n$. The ground-state entanglement entropy is then calculated by tracing out the $N-n$ chains. After employing a set of canonical transformations, one can show that the reduced density matrix $\rho_A$ (or $\rho(z_1,\cdots, z_n;z_1',\cdots, z_n')$ in terms of transformed coordinates) is the product of density matrices of two coupled oscillators. This simplifies the calculation substantially, as the entanglement entropy can be expressed in terms of the eigenvalues of the density matrices of two coupled oscillators. As noted by Srednicki, the decomposition of the chain actually introduces a thermal nature for the reduced density matrix. Since there is now a boundary between traced-out modes and the others, a source of mode-mixing is provided by the boundary. This effectively cuts the long-wavelength modes
and creates an effective horizon like black holes. The reduced density matrix $\rho_A$ now reads –  
\begin{equation}
    \rho_A=\otimes_{i=1}^{n} \rho_i \label{rho_dc},
\end{equation}
with each $\rho_i$ becoming the density matrix of a thermal oscillator. 
In diagonal coordinates using ground state wave functions $\psi_0$, the reduced density matrix for a coupled oscillator can be obtained by integrating out the second oscillator's degree of freedom enabling $\rho_A=Tr_B \rho_{AB}$.
\begin{equation}
    \rho_A(z_1, z_1')=\int_{-\infty}^{\infty}\psi_0(z_1,z_2)\psi^*_0(z_2,z_1')dz_2,
\end{equation}
where \begin{equation}
    \psi_0(z_1,z_2)=\pi^{-1/2}(\omega_1\omega_2)^{1/4}exp[-(\omega_1 z_1^2+\omega_2 z_2^2)],
\end{equation}
with $z_1=\frac{(x_1\pm x_2)}{\sqrt{2}}$; and $\omega_1=\omega;\, \omega_2=\sqrt{\omega^2+2k}.$ Upon integration the reduced density matrix in terms of the original coordinates becomes-
\begin{equation}
\rho_A(x_1,x_1')=    \pi^{- \frac{1}{2}}
\left(
\frac{2\omega_{1}\omega_{2}}
{\omega_{1}+\omega_{2}}
\right)^{\frac{1}{2}}
\exp\left[
-
\frac{(\omega_{1}+\omega_{2})^{2}+2\omega_{1}\omega_{2}}
{4(\omega_{1}+\omega_{2})}
\frac{x_{1}^{2}+x_{1}^{\prime 2}}{2}
+
\frac{(\omega_{1}-\omega_{2})^{2}}
{4(\omega_{1}+\omega_{2})}
x_{1}x_{1}^{\prime}
\right]
    \end{equation}

The eigenvalues of a single coupled oscillator can be obtained either in the Fock basis or in the position basis. Using the equation, \begin{equation}
    \int_{-\infty}^{\infty} \rho_A(x, x') 
    \phi_n (x') dx'=p_n \phi_n(x),
\end{equation} one obtains the eigenvalues $p_n$. It turns out to be:
\begin{equation}\label{proba}
 p_n= (1-\xi^2)\xi^{2n},
\end{equation}
where \[\xi =
\frac{
\omega_{2}^{1/2} - \omega_{1}^{1/2}
}{
\omega_{2}^{1/2} + \omega_{1}^{1/2}
}.\]

The eigenfunctions are Hermite polynomials
\[
\phi_n(x) = H_n\!\left(\omega^{1/2} x\right)
\exp\!\left(-\frac{\omega x^2}{2}\right),
\] with $\omega=(\omega_1 \omega_2)^{1/2}.$ The entanglement entropy (EE) is now expressed as the von-Neumann entropy of the subsystem $S=-\sum_m p_m\log p_m$ yielding 

\begin{equation}\label{EE1}
    S_{EE}=\sum_{i} S_i=-\log(1-\xi^2)-\frac{\xi^2}{1-\xi^2}\log \xi^2.
\end{equation}

We next focus our attention to the IHO. The Hamiltonian for an IHO can be expressed as:

\begin{equation}\label{H-niho}
H = \frac{1}{2} \sum_{i=1}^{N} p_i^2 
- \frac{1}{2} \sum_{i,j=1}^{N} x_i K_{ij} x_j
\end{equation}
where $K_{ij}$ is a positive-definite symmetric matrix. We continue to restrict ourselves to nearest-neighbour interactions. 

The density operator for a single inverted oscillator can be defined in terms of the generalised eigenvectors as: $\rho=|f_n^+\rangle\langle f_n^-|$ or $\rho(x,x')=\langle x|f_n^+\rangle\langle f_n^-|x'\rangle  $.
Using canonical transformations very much similar to what one deploys for HO, we can easily find the zeroth state wavefunction (decay state) for a coupled oscillator $(N=2)$ using \eqref{Hermite}:
\begin{align}
    f_0^+(x_1,x_2)= C \exp\Bigg\{
&-\frac{i}{4}
\left[
\sqrt{\omega_0^2+k}
+\sqrt{\omega_0^2-k}
\right](x_1^2+x_2^2)
\nonumber \\[6pt]
&-\frac{i}{2}
\left[
\sqrt{\omega_0^2+k}
-\sqrt{\omega_0^2-k}
\right]x_1x_2
\Bigg\},
\label{2cwf}\end{align}
where $C$ is the normalisation constant.

The normalisation of these zeroth-state energy eigenfunctions are to be performed using a test function that is rapidly converging to zero at the asymptotic boundaries. Next, we calculate the zeroth state density matrix by tracing over one IHO d.o.f. We use the resonant states again and employ the tempered function
\begin{equation}
\phi(x)=x^m e^{-\alpha^2x^2/2},
\qquad
m\in\mathbb{Z}_{\geq0},\;
\alpha\in\mathbb{R},
\end{equation}
which belongs to the Schwartz space \(S(\mathbb{R}_x)\). We use the following Gauss--Fresnel integral, 
\begin{equation}
\int_{-\infty}^{\infty}
\left(x^m e^{-\alpha^2x^2/2}\right)^*
x^n
e^{\pm i\beta^2x^2/2}\,dx
=
(m+n-1)!!
\sqrt{\frac{2\pi}
{\left(\alpha^2\mp i\beta^2\right)^{\,m+n+1}}},
\end{equation}
provided that \(m+n\) is even, and zero
otherwise. The reduced density matrix now reads:
\begin{equation}\label{dm_ciho}
    \rho(x_1, x_1')=\int_{-\infty}^{\infty}f_0^-(x_1,x_2)f_0^+(x_2,x_1') e^{-\alpha^2 (x_1^2+2 x_2^2+x_1'^2)/2} dx_2.
\end{equation}
The integral in \eqref{dm_ciho} can be done
using the standard Gaussian integration techniques. Using the generalized zeroth-state wavefunction for two coupled inverted harmonic oscillators, we define
\[
\Omega_\pm=\sqrt{\omega_0^2\pm k}.
\]

The wavefunction then reads,
\[
f_0(x_1,x_2)
=
C
\exp\left[
-\frac{i}{4}
\left(
\Omega_+(x_1+x_2)^2
+
\Omega_-(x_1-x_2)^2
\right)
\right].
\]

We can write the product 
\[
f_0^-(x_1,x_2)f_0^+(x_2,x_1')
=
|C|^2
e^{ia(x_1^2-x_1'^2)}
e^{ib(x_1-x_1')x_2},
\]
where
\[
a=\frac{\Omega_++\Omega_-}{4},
\qquad
b=\frac{\Omega_+-\Omega_-}{2}.
\]

Substituting into the integral gives
\begin{align}
\rho(x_1,x_1')
={}&|C|^2
e^{-\frac{\alpha^2}{2}(x_1^2+x_1'^2)}
e^{ia(x_1^2-x_1'^2)}
\nonumber\\
&\times
\int_{-\infty}^{\infty}
e^{-\alpha^2x_2^2+ib(x_1-x_1')x_2}\,dx_2.
\end{align}
Using the Gaussian integral
\[
\int_{-\infty}^{\infty}e^{-Ax^2+Bx}\,dx
=
\sqrt{\frac{\pi}{A}}\,
e^{B^2/(4A)},
\qquad \Re(A)>0,
\]
we obtain
\begin{align}\label{rhonew1}
\rho(x_1,x_1')
={}&\frac{|C|^2\sqrt{\pi}}{\alpha}
\exp\Bigg[
-\frac{\alpha^2}{2}(x_1^2+x_1'^2)
+i\frac{\Omega_++\Omega_-}{4}
(x_1^2-x_1'^2)
\nonumber \\[4pt]
&\qquad
-\frac{(\Omega_+-\Omega_-)^2}{16\alpha^2}
(x_1-x_1')^2
\Bigg].
\end{align}

It is to be noted, without the Gaussian damping terms, the integral reduces to a constant value indicating a situation where the trace of density matrix fails to be well defined. Next, we construct the eigenvalues of the reduced density matrix that are needed to find the bipartite entanglement entropy. The desired equation is
\begin{equation}
    \int_{-\infty}^{\infty} \rho(x, x') 
    \phi_n (x') dx'=\lambda_n \phi_n(x),
\end{equation}
where $\phi_n$ denotes the eigenfunctions and $\lambda_n$ the eigenvalues. Since every term in the exponent is either a real or complex Gaussian, we can get the desired eigenfunctions. 

Motivated by the Gaussian structure of the kernel, we assume the zeroth eigenfunction to be of the form
\begin{equation}\label{trial_gs}
\phi_{0}(x)
=
N_c
\exp\left(-\frac{\beta}{2}x^{2}\right),
\end{equation}
where the width parameter \(\beta\) is allowed to be complex, subject to the condition \(\Re(\beta)>0\). The details of obtaining the zeroth-state eigenfunction and the eigenvalue are given in the Appendix A.

The zeroth state analysis shows (see \eqref{labda0}, \eqref{lambdabasic}) that the eigenfunctions are complex Gaussians and can be represented by Hermite polynomials., 
\begin{equation}
    \phi_{0}(x)
=N_c
\exp
\left[\left(
-\frac{\alpha^2}{2}
+i\frac{\Omega_++\Omega_-}{4}
-\frac{(\Omega_+-\Omega_-)^2}{16\alpha^2}
+\frac{N^2}{4M}
\right)x^2\right].
\end{equation}

The higher-order states can be obtained by setting $\lambda_n=\lambda_0 \xi^n$, where 
\begin{align}
\xi=
\frac{D}{s+\sqrt{s^2-D^2}}.
\end{align}
Where,
\[
s=\frac{\alpha^2}{2}+D
=\frac{\alpha^2}{2}
+\frac{\left(\sqrt{\omega_0^2+k}-\sqrt{\omega_0^2-k}\right)^2}
{16\alpha^2}
,\]
\[
D=\frac{\left(\sqrt{\omega_0^2+k}-\sqrt{\omega_0^2-k}\right)^2}
{16\alpha^2}
.\]
The calculation of entanglement entropy $S_{EE}=- Tr \lambda_n \log \lambda_n,$ may now proceed in a similar manner as has been done for the HO case. 
What is the interpretation of the parameter $\alpha?$  Recall we introduced the parameter $\alpha$ through the rapidly decaying function in the expression for the zeroth-state energy for the coupled inverted oscillators model. It is not unusual to have some regulator or arbitrary parameter present in the EE that are not present in the original Hamiltonian. EE may depend on the subsystem size, UV cut-off etc \cite{PasqualeCalabrese_2004, Srednicki:1993im}.  It is imperative that this $\alpha$ parameter here plays the role of a cut-off in the position space, as the zeroth state is non-renormalizable. This is reminiscent of the box normalization of a free particle wave function \cite{Yuce_2006}. However, we may consider the reduced density matrix and its eigenvalues in the limit $\alpha \to 0$ to gain insights about the system's behaviour in the absence of this parameter. The most important parameter is $\beta$, which in the limit mentioned (assuming $\alpha >0)$ reduces to

\begin{align}\label{betalimit}
    \lim_{\alpha\to0^+}\beta
=
-\frac{i}{2}\left(
\sqrt{\omega_0^2+k}+\sqrt{\omega_0^2-k}
\right).
\end{align}

Further, \begin{align}\label{lamlimit}
\lim_{\alpha\to0^+}\lambda_0
=
\frac{4\pi|C|^2}
{\left|\sqrt{\omega_0^2+k}-\sqrt{\omega_0^2-k}\right|}.
\end{align} This assumes fixed $k\neq0, |k|<\omega_0^2$, and the reduced kernel normalisation used above. The trace of the reduced density matrix in this limit reads \[
\operatorname{Tr}\rho_\alpha
=\frac{\pi|C|^2}{\alpha^2}.
\] So the normalization requires \[
|C|^2=\frac{\alpha^2}{\pi}, 
\] 

For a consistency check, we let both, $k$ and $\alpha$ tend to zero; we get 
\begin{align}\lim_{\alpha\to0^+}\beta=-i\omega_0.\end{align}  Therefore the zeroth eigenfunction reduces to the desired form  $(\sim e^{\frac{i\omega_0 x^2}{2}}).$ The order of the limits should be taken carefully, as mentioned above; otherwise, the limits will not produce any acceptable answer. This is because the position space density matrices are not well defined in the limit $\alpha \to 0$ alone. Clearly, $\alpha$ acts as a regulator of the IHO states, and determines the decay (or growth) rates of the resonant states. Because of the inherent instability and the need to regularise the integrals, obtaining a closed expression for $S_{EE}$ for N-coupled IHOs is not easy. In the next section, we formally derive an explicit expression for EE for coupled IHOs using operator techniques.


\section{Entanglement entropy for IHO chain using Ladder operators}

We begin this section by reviewing the ladder operator formalism for an inverted harmonic oscillator. One can formally define generalised ladder operators following the techniques used for standard HO:
\begin{equation}\label{caiho}
b^\pm
=
\frac{1}{\sqrt{2 \omega_{io}}}
\left(
\omega_{io} x\pm p
\right),
\end{equation}
which satisfy the commutation relations
\begin{equation}\label{commut}
[b^+,b^-]=-i; ~~[b^+,b^+]=0; ~~[b^-,b^-]=0.
\end{equation}

The Hamiltonian can then be rewritten as
\begin{equation}\label{Hiho}
H_{\rm IHO}
=
-\frac{\omega_{io}}{2}
(b^+b^-+b^-b^+).
\end{equation}

Generalized vacuum states are defined by \cite{2000NCimB.115..325S}
\begin{equation}
b^\mp |f_0^\pm\rangle =0.
\end{equation}

Excited states are generated through repeated action of the ladder operators:
\begin{equation}
|f_n^\pm\rangle \propto  (b^\pm)^n |f_0^\pm\rangle.
\end{equation} 
An analogue of number operator can be defined as \begin{equation}
    \hat{N}=\frac{1}{2}
(b^+b^-+b^-b^+). 
\end{equation}
Using a series of commutation relations like
\begin{equation}\label{commut2}
[\hat{N},b^\pm]=\pm ib^\pm; ~~[b^\mp,(b^\pm)^n]=\pm in(b^\pm)^{n-1},
\end{equation}
one can show that the Hamiltonian  possesses imaginary energy eigenvalues: \[H_{IHO} |f_n^\pm\rangle=\pm E_n |f_n^\pm\rangle,\]
with 
\begin{equation}
E_n=\mp i\omega_{io}\left(n+\frac12\right),\,\, n=0,1,2.\cdots
\end{equation}
The energy states describe exponentially growing or decaying modes (Gamow states). In a more careful treatment, one needs to promote the operators into $S^\times(\mathbb{R})$, but it can be shown that the results remain the same \cite{2000PhLA..273...37S}.

We now obtain the entanglement entropy of N-coupled IHO using the ladder operators and the eigenstates of the Hamiltonian $H_{IHO}$. We recall that the density operator for the $i^{th}$ HO can be expressed as 
\begin{equation}\label{rhobasic}
    \rho_i=\frac{1}{Z_i}e^{-\omega_i(a^{\dagger}_ia_i+\frac{1}{2})/K_BT_i}=\frac{1}{Z_i}e^{-\beta_i H_i},
\end{equation}
where, $\epsilon_i$ is related to the frequency of the $i^{th}$ oscillator $\omega_i/k_BT_i.$ $Z_i$ is the partition function.  The ladder operators in Eq. (\ref{rhobasic}) must be understood as the ones that reduce the covariance matrix in its normal form. In the biorthogonal bases of IHO, the partition function reads,
\begin{equation}\label{parfn}
    Z_i^{\pm}= \sum_{n_i=0}^\infty \langle f_{n_i}^{\pm}|e^{-\beta_i H_{IHO}^i}|f_{n_i}^{\mp}\rangle=\pm\frac{e^{\frac{i \beta_i \omega_i}{2}}}{e^{i\beta_i \omega_i}-1},
\end{equation} one yields the following density matrix\footnote{Here the frequencies of IHO are denoted by $\omega_i.$}.
\begin{equation}\label{rhoih}
    \rho_i^{\pm}=\pm(e^{i \beta_i \omega_i}-1)e^{-i \beta_i \omega_i/2 - \beta H_i}|f_{n_i}^{\mp}\rangle\langle f_{n_i}^{\pm}|.
\end{equation}
Where, for brevity, we have discarded the subscript $IHO$ from the IHO Hamiltonian. Both the density matrices obtained by choosing the two possible combinations of biorthogonal bases  obey the $Tr \rho=1$ property. Next, the average occupation number is obtained using the standard formula (dropping the superscript $'-'$ on $\rho_i)$

\begin{equation}\label{occupn}
    <n_i>=Tr (\rho_i \hat{N_i})=-\frac{i \left(e^{\omega_i  \text{i$\beta_i $}}+1\right)}{2 \left(e^{\omega_i  \text{i$\beta_i $}}-1\right)}=-\frac{i}{2}\cot{\frac{\beta_i \omega_i}{2}}
\end{equation}
On the other hand, by direct calculation  we get 
\begin{align}<x_i^2>=Tr(\rho \hat{2N})=\frac{i}{2\omega_i}\cot{\frac{\beta_i \omega_i}{2}};~~<p_i^2>=\frac{ i\omega_i}{2}\cot{\frac{\beta_i \omega_i}{2}}.\end{align}

Here we have used $<f_n^+|b^+b^+|f_n^->=0$ since, $<f_n^+|b^+ \propto \overline{b^-|f_n^->}= <f_{n-1}^+| .$ It has to be noted that the Hamiltonian for IHO is proportional to $\hat{N}$, whereas for HO it is $\hat{N}+1$. 

Next, we can cast the density matrix $\rho_i$ in terms of $<n_i>$ as
\begin{equation}
\rho_i= {1\over{(<n'_i>-1)}}\left(2\frac{<n'_i>+1}{<n'_i>-1}\right)^{n_i}.
\end{equation}
Here, we considered that the absolute value of the mean occupation number and defined \[<n'_i>=|<n_i>|.\]
The generalized entanglement entropy $S_{EE}$ for $i^{th}$ oscillator is given by 
\begin{equation}\label{EE_f}
-Tr(\rho_i\log \rho_i)
=<f_{n_i}^+|\sum_{n_i=0}^\infty\frac{1}{<n'>+1}\left(2\frac{<n'_i>+1}{<n'_i>-1}\right)^{n_i}
\log\left[\frac{1}{<n'>+1}\left(2\frac{<n'_i>+1}{<n'_i>-1}\right)^{n_i}\right]|f_{n_i}^->
. 
\end{equation}
After evaluating the following two infinite sums 
\[\sum_{n_i=0}^\infty\left(2\frac{<n'_i>+1}{<n'_i>-1}\right)^{n_i}=\frac{1-<n'_i>}{3+<n'_i>}\,;\,\sum_{n_i=0}^\infty n_i\left(2\frac{<n'_i>+1}{<n'_i>-1}\right)^{n_i}=\frac{2(1-<n'_i>^2)}{(3+<n'_i>)^2},\]
we obtain 
\begin{equation}
    \label{finalEE}
    S^{EE}_i=-\frac{<n'_i>-1}{(<n'_i>+1)(<n'_i>+3)}\log(<n'_i>+1)-\frac{2(<n'_i>-1)}{(<n'_i>+3)^2}\log\left(2\frac{<n'_i>+1}{<n'_i>-1}\right).
\end{equation}
It must be mentioned that the infinite series presented above are only convergent within a limited range, and that forces the number $<n_i'>$ to be within $-3$ and $-{1\over3}.$ This negative average occupancy indicates the unstable nature of the system. A negative occupation number can occur in some particular situations where the system has quasi-hermiticity, pseudo-hermiticity, etc. Here, the introduction of a biorthogonal basis together with the apparent relation of the IHO frequency to the HO frequency in the energy eigenspectra through an analytical continuation invites such a situation. If, instead of the absolute value, we want to replace the actual mean occupation number in the expression \eqref{finalEE} to obtain the EE naively, the infinite sums will not converge in general.  A complex EE indicates the quasi-hermiticity of the system. Our absolute value prescription avoids this issue. With the individual EE for a coupled IHO on hand, the total generalized entanglement entropy of the subsystem can be obtained as
\begin{equation}
    S_A=\sum_{i=1}^n S^{EE}_i.
\end{equation}

The relation between this treatment based on the Fock basis and the treatment outlined in section II, can be obtained by relating the parameter $\xi$ in \eqref{proba}, with the probabilistic concept of density operator \eqref{rhoih} (using $\rho^-_i$)
\[(1-\xi^2)\xi^{2n}=(1-e^{i \beta_i \omega_i}) e^{i \beta_i \omega_i n} .\]
This relation identifies $\xi=e^{i\beta \omega/2}.$ The system's imaginary frequencies are related to the temperature. As explained earlier the thermality is introduced by tracing out subsystem's degrees of freedom. 
\section{Conclusion}
We have calculated the entanglement entropy for a system of coupled inverted harmonic oscillators. IHOs provide the simplest unstable and quasi-hermitian system that captures many features of non-hermitian systems, like biorthogonal bases, complex density matrix elements, etc. We used the position basis wavefunctions to demonstrate the two-coupled IHOs and indicated the subtle issues of obtaining the entanglement entropy for systems having complex energy eigenvalues or eigenfunctions. A closed expression for EE for a subsystem of N-coupled IHOs has been obtained by working on the eigenvectors of the energy or the Hamiltonian in terms of ladder operators. The result shows that the EE can not naively be obtained by replacing $\omega\to i\omega_0$ in the von-Neumann entropy of the corresponding chain of HO without introducing a regulator. We have proposed an alternative path by extracting the absolute value of the density operator and casting the EE as a real quantity. 

Although the absolute value prescription yields a real EE, but an alternative expression for EE may still be obtained without introducing this prescription. This can be obtained either by introducing a regulator to the divergent infinite sums, or interpreting the result as an analytic continued version of the \eqref{finalEE}. In this scenario, the EE becomes complex. Emergence of an imaginary part of entropy due to the complex energy eigenvalues for IHO has been considered in \cite{PhysRevE.63.056101}, where a statistical mechanics description of such behaviour is proposed. However, a completely acceptable information-theoretic description might still have to be developed. The complex nature of EE may have a deep connection with non-unitary field theories \cite{n578-ljd5}.  It will be interesting to see if the study could also be extended to many-body physics or quantum fields using a generalized Fock basis. The relation of such unstable, non-Hermitian systems to the Ginibre ensemble framework and the corresponding Page curve-like dynamics may be another important direction. It will also be interesting to explore any relation to the complex-valued pseudo entropy appearing in the holographic EE in de Sitter space \cite{Doi:2022iyj} or in non-unitary dynamics \cite{Cipolloni:2022fej}.  \\

\textbf{Acknowledgements:} The author acknowledges initial discussions with S. Naga Sai Srinivas. The author thanks useful interactions with Arpan Bhattacharyya and Koushik Ray. A part of the work has been presented at the workshop on Frontiers in Quantum Information \& Technologies held between 26-29 March 2026 at IIIT Allahabad. It was funded by ANRF through the grant: SSY/2026/000182.

\appendix
\section{Derivation of position space kernel for a coupled IHO}

From \eqref{rhonew1} we rewrite the kernel as,
\begin{equation}\label{eveqrho}
\rho(x,x')\phi_0(x')
=
\frac{|C|^2 \sqrt{\pi}N_c}{\alpha}
\exp\left[\left(
-\frac{\alpha^2}{2}+i\frac{\Omega_++\Omega_-}{4}
-\frac{(\Omega_+-\Omega_-)^2}{16\alpha^2}
\right)x^2\right] \exp\left[-M x'^2+Nx x'\right],
\end{equation}
where

\begin{equation}\label{MN}
M
=
\left(
-\frac{\alpha^2+\beta}{2}+i\frac{\Omega_++\Omega_-}{4}
-\frac{(\Omega_+-\Omega_-)^2}{16\alpha^2}
\right),\, N= \frac{(\Omega_+-\Omega_-)^2}{8\alpha^2}
.\end{equation}
Using the Gaussian-Fresnel integral
\begin{equation}
\int_{-\infty}^{\infty}
e^{-Mx'^{2}+Nxx'}\,dx'
=
\sqrt{\frac{\pi}{M}}
\exp\left(
\frac{N^{2}x^{2}}{4M}
\right),
\qquad
\Re(M)>0,
\end{equation}

we get,
\begin{align}\label{labda0}
\lambda_{0}\phi_{0}(x)
={}&
\frac{|C|^2\pi N_c}{\alpha\sqrt{M}}
\times
\exp
\left[\left(
-\frac{\alpha^2}{2}
+i\frac{\Omega_++\Omega_-}{4}
-\frac{(\Omega_+-\Omega_-)^2}{16\alpha^2}
+\frac{N^2}{4M}
\right)x^2\right]
.
\end{align}

Since the right-hand side must reproduce the Gaussian form of $\phi_0(x)$, comparison of the quadratic coefficients of both sides yields
\begin{equation}
-\frac{\beta}{2}
=
\frac{\alpha^2}{2}
+i\frac{\Omega_++\Omega_-}{4}
-\frac{(\Omega_+-\Omega_-)^2}{16\alpha^2}
+\frac{N^2}{4M},
\end{equation}

Consequently,\begin{equation}
\beta
=
-\alpha^2
-i\frac{\Omega_++\Omega_-}{2}
+\frac{(\Omega_+-\Omega_-)^2}{8\alpha^2}
-\frac{N^2}{2M}
.\end{equation} 

Using the explicit forms of $M, N$ one get $\beta$ as,

\begin{align}\label{beta}
\beta(\alpha,k)
={}&-\alpha^2
-\frac{i}{2}
\left(
\sqrt{\omega_0^2+k}
+\sqrt{\omega_0^2-k}
\right)
\nonumber\\[6pt]
&+\frac{
\left(
\sqrt{\omega_0^2+k}
-\sqrt{\omega_0^2-k}
\right)^2
}{8\alpha^2}
\nonumber\\[6pt]
&-\frac{
\left(
\sqrt{\omega_0^2+k}
-\sqrt{\omega_0^2-k}
\right)^4
}{
128\alpha^4
\left[
\displaystyle
\frac{\alpha^2}{2}
+\frac{i}{4}
\left(
\sqrt{\omega_0^2+k}
+\sqrt{\omega_0^2-k}
\right)
+\frac{
\left(
\sqrt{\omega_0^2+k}
-\sqrt{\omega_0^2-k}
\right)^2
}{16\alpha^2}
\right]
}.
\end{align}

Finally, the corresponding zeroth-state eigenvalue is
\begin{equation}\label{lambdabasic}
    \lambda_0
=
\frac{|C|^2\pi}{\alpha\sqrt {M}}.
\end{equation}

The full expression for $\lambda_0$ is as follows:
\begin{equation}\label{lamfull}
\lambda_0
=
\frac{|C|^2\pi}{
\alpha\left(
\displaystyle
\frac{\alpha^2}{2}
+\frac{i}{4}
\left(
\sqrt{\omega_0^2+g}
+\sqrt{\omega_0^2-g}
\right)
+\frac{
\left(
\sqrt{\omega_0^2+g}
-\sqrt{\omega_0^2-g}
\right)^2
}{16\alpha^2}
\right)^{1/2}}.
\end{equation}

\bibliographystyle{apsrev}
\bibliography{ref}
\end{document}